\documentclass[12pt,psf,epsf]{JHEP3}
\usepackage{euscript,amsmath,amssymb,epsfig,latexsym}

\usepackage{amsmath,amsfonts,amssymb}
\usepackage{graphicx}

\newcommand{\be}{\begin{equation}}
\newcommand{\ee}{\end{equation}}
\newcommand{\bea}{\begin{eqnarray}}
\newcommand{\eea}{\end{eqnarray}}

\newcommand{\arctanh}{\mbox{arctanh}}

\title{On Holographic Thermalization and Dethermalization
of 
Quark-Gluon Plasma}
\author{I.Ya. Arefeva and I.V.Volovich\\
\small{\it Steklov Mathematical Institute, RAS, Moscow}}
\date {\today}
\maketitle

\abstract{
 We estimate   the ratio of the thermalization time over the
freeze-out time using a holographic AdS-Vaidya model for quark-gluon
plasma formed in the heavy ion collisions.
 In the  model the process of thermalization  is
described as formation of the black hole in AdS space while
dethermalization process, related  with the freeze-out, as   the
black hole evaporation   due to the Hawking radiation that
 is modeled by the  Vaidya metric with a negative mass. In this model the
thermalization takes place only at small scales and absent in the
infrared region. At small scales the system tends towards a state of the
thermal equilibrium only for the short time after which the processes
of dethermalization starts. In this simple model the
dethermalization time  has a low bound about 7 fm/c which is consistent with experimental data. 
}

\begin{document}
\newpage
\section{Introduction}

 In the process of collision of two ions one observes a
rapid local thermalization of the system of partons with further
expansion and cooling of system which leads to hadronization and
multiple production of particles. Experiments indicate to a very
short thermalization time, $\tau _{therm}\sim 1$ fm/c for the
quark-gluon plasma(QGP) formed in heavy ion collisions,  while
the freeze-out time is of order 20 fm/c
 \cite{rhic,Gyulassy:2004zy,fluid2,Iancu:2012xa,Gelis,Muller11}.

In more details, the time schedule of the process is the  following.
Up to a time $\sim 0.02$ fm/c $\,$"hard" $\,$ processes take place and
they are responsible for $\,$"hard"$\,$ particles, which can be observed at
detectors. Up to  a time $\sim 0.2$ fm/c $\,$"semi-hard"$\,$ processes
take place and they produce  the most of the
 "multiplicity"$\,$  in the
final state. Then
at the thermalization time of  order  $\tau_{therm}\sim 1$ fm/c
the system reaches a local thermal
equilibrium state, called QGP. After that the  evolution of QGP is described by equations
of hydrodynamics and after the time of  order $\tau_{hadr}\sim 10$
fm/c, when due to the separation of the  colliding ions the
temperature becomes lower than the deconfinement temperature, a hot
hadron gas is formed.  Upon the further expansion and cooling, around
the freeze-out time, we refer to this time as the dethermalization time  $\tau_{det}\sim 20$ fm/c,  the
density of the hadron gas become sufficiently low and the system
decays into free hadrons, which can be observed at detectors. Therefore in
experiments on heavy ion collisions there is the following hierarchy
of time scales:
\be \label{hiehad}
\tau_{therm}<\tau_{hydro}<\tau_{hadr}<\tau_{det}. \ee

 There are many attempts to describe   the
process of heavy ion collisions and QGP formation in the QCD framework.
One of difficulties is that one has   to compute the time depended correlation functions
in the strong coupling regime since the QGP is a strong coupling
system \cite{Gyulassy:2004zy}.  In the recent years a
powerful  approach to these problems is pursued which is based on
 a holographic duality between the strong coupling
quantum field in $d$-dimensional Minkowski space and classical
gravity in $d+1$-dimensional anti-de Sitter space (AdS) \cite{Malda,GKP,Witten}.
In particular, there is a considerable progress in
 the holographic description of equilibrium  QGP  \cite{EQ-QGP}.

Holographic thermalization describes the
thermalization process  as a
process of formation of a black hole in AdS. There are several scenarios
to produce a black hole in AdS \cite{Gubser}-\cite{Elena}.
One of them uses the AdS-Vaidya metric \cite{Keski,Hubeny:2007xt,AbajoArrastia:2010yt,Balasubramanian:2012,
Balasubramanian:2011ur,Callan:2012ip}.

In this paper we consider an influence of the Hawking radiation on the
thermalization process. We describe the
evaporation of black hole due to the Hawking radiation by the
Vaidya metric with negative mass. We use a special form of the AdS-Vaidya metric, see eq. (\ref{mv-pm-m}) below,
 to describe thermalization and subsequent dethermalization. We show that the evaporation
of the black hole in AdS leads to an interesting phenomena in the $d$-dimensional Minkowski space-time
-- thermalization is possible only at small distances and impossible
in the infrared region.

The Hawking radiation from a  Vaidya black hole was considered in
\cite{VZF} where it was shown that allowance for the nonstationarity reduces
to allowing for the dependence of the temperature of the Hawking radiation
on the retarded time.
Note that the Hawking radiation in holographic approach
was also discussed in \cite{Chesler:2011ds}.

The paper is organized as follows.
In Sect. 2 we remind the basic formulas related with the AdS-Vaidya
metric which are used for the holographic thermalization
\cite{Balasubramanian:2012}-\cite{Callan:2012ip}. Sect.3 discusses the AdS-Vaidya metric
with negative mass to model the Hawking radiation. We show that
thermalization for the two-point correlation functions holds only at
small distances. In Sect. 4 we discuss attempts to fit  the time hierarchy (\ref{hiehad})
 obtained in Pb+Pb collisions  at  LHC and in Au+Au collisions  at RIHC,  by  a special form of the AdS-Vaidya metric. In Appendix  we present an explicit formula for a non-equal time two points correlation function for the dethermalization process.  This formula is a counterpart of non-equal time two points correlation functions for the thermalization process found in \cite{Balasubramanian:2012}-\cite{Callan:2012ip}.

\section {Holographic Thermalization with AdS-Vadiya metric}

Holographic duality prescribes that the vacuum correlation functions
in quantum field theory in the strong coupling regime in
$4$-dimensional Minkowski space can be computed by using the action
functional for corresponding fields in   $d+1$-dimensional AdS
\cite{Malda,GKP,Witten}.  Thermal states also admit a holographic
description by using a black hole AdS metric \cite{Witten,Son}.
The holographic description of equilibrium  QGP  was tested
in numerical works, see the review paper \cite{EQ-QGP} and refs therein.

Hypothesis on holographic thermalization asserts  that the
thermalization process can be described in the dual framework as the
process of formation of a black hole in AdS. The process of formation of a black hole
can be initiated by a perturbation of the initial AdS metric.
Formations of black holes in gravitational collapse or in the
collision of ultra-relativistic particles are traditional difficult
problems in general relativity, see for example \cite{IA-cat} and refs therein.
In
AdS there are specific features  related with the compression of
AdS metric. In the context of holographic description of heavy-ions collisions
formation of black holes in  collision of gravitational waves
in AdS  has been considered in numerous works
\cite{Gubser,Chesler:2008hg,Alvarez,Lin:2009pn,Yaffe,ABGJ,KirTalio11,ABP}.
Gravitational collapse occurring as a result of the weak perturbation in AdS is
considered in
  \cite{Bhattacharyya:2009uu}, where  it is shown that the system
  evolves to the state of thermal equilibrium almost instantly.
  This result stimulates application of the AdS-Vaidya metric for the thin shell
  in the holographic description of thermalization
  \cite{Keski,Hubeny:2007xt,AbajoArrastia:2010yt,Balasubramanian:2012,Balasubramanian:2011ur,Callan:2012ip}.

The ($d+1$)-dimensional infalling matter shell in AdS in
Poincar\'e coordinates is described  by the Vaidya metric

\be ds^2 = \frac{1}{z^2}\left[-\left(1 - m(v)z^d\right) dv^2 - 2
dz\, dv + d\mathbf{x}^2 \right] \,, \label{eq:Vaidya} \ee
where $v$
is the null coordinate,
 $\mathbf{x}=(x_1,\dots,x_{d-1})$ are the spatial
coordinates on the boundary $z=0$ and we have set the AdS
radius equal to 1.
We take  $m(v)$ in the form
 \be \label{mv}
m(v)=M\theta(v), \ee
where  $M$ is a constant and  $\theta(v)$ is the Heaviside function.

For  $m(v) =M$  the change of variables
\be dv=dt-\frac{dz}{1- M\, z^d} \label{eq:dt_and_dv} \ee
 brings  (\ref{eq:Vaidya}) to the standard metric of the black hole in
AdS in the Poincare coordinates
 \be ds^2 = \frac{1}{z^2} \left[-\left(1 -
M z^d\right) dt^2 + \frac{ dz^2}{1-M z^d} + d\mathbf{x}^2 \right]\,.
\ee
For $v<0$ the metric (\ref{eq:Vaidya}) with $m(v)$ in the form (\ref{mv})
is just the  AdS metric.

 The holographic prescription permits to calculate an
 equal-time two-point function of operators ${\cal O}(x,t)$
 with large conformal dimension $\Delta$ in
 dual theory of gravity by using the geodesic approximation
\cite{Balasubramanian:1999zv}:
 \be
 \langle{\cal
O}(\mathbf{x},t){\cal O}(0,t)\rangle \sim \exp[ - \Delta\, {\cal
L}(\mathbf{x},t) ].\label{cor}
\ee
Here ${\cal L}(\mathbf{x},t)$ is the length of the geodesic
that begins and ends at the boundary points $(0,t)$ and $(x,t)$.
The vacuum  correlation functions correspond to the computation of
geodesics in AdS,  thermal ones in AdS with a black hole
($BHAdS$) while correlation functions describing the process of thermalization
correspond to geodesics in the AdS-Vadiya (\ref{eq:Vaidya}) with (\ref{mv}).
 These geodesics are studied in papers
\cite{Hubeny:2007xt,AbajoArrastia:2010yt, Balasubramanian:2012,
Balasubramanian:2011ur,Callan:2012ip}.
These considerations give an estimation of the thermalization time
for two equal time points at the boundary. In particular, for $d=2$
the thermalization time is equal to the half of the space distance between
these points.
For $d>2$ the thermalization time decreases. The Vaidya Reissner-Nordstrom AdS metric
increases the thermalization time in all dimensions \cite{1205.1548,Elena}

\section{The Hawking Radiation and Dethermalization}
\begin{figure}[h!]
\begin{center}
\includegraphics[width=30mm]{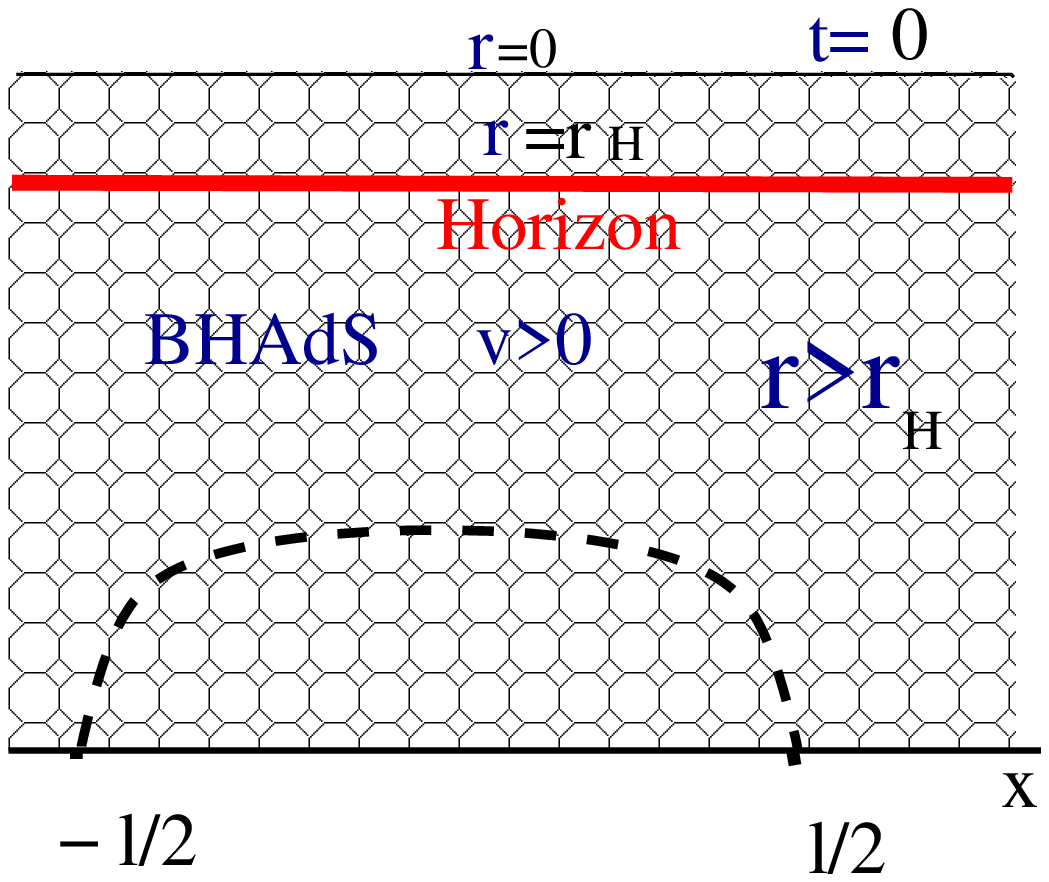}A.\,\,\,\,\,
\includegraphics[width=30mm]{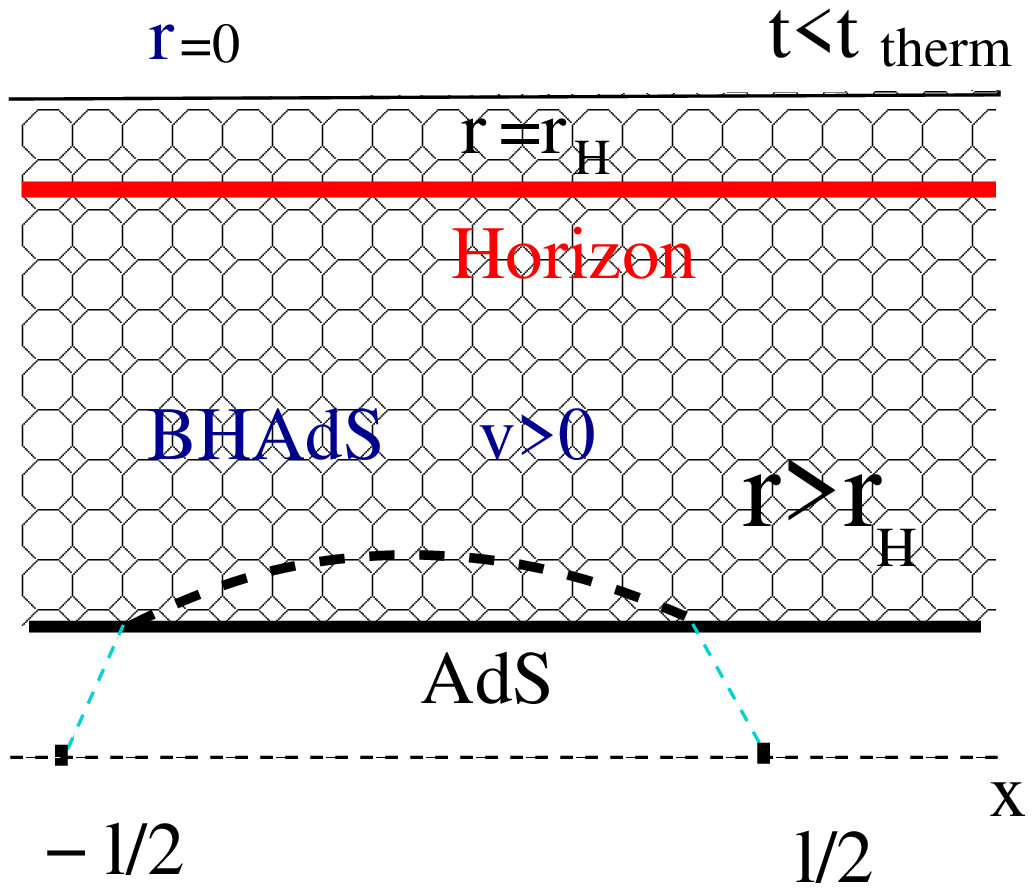}B.\,\,\,\,\,
\includegraphics[width=30mm]{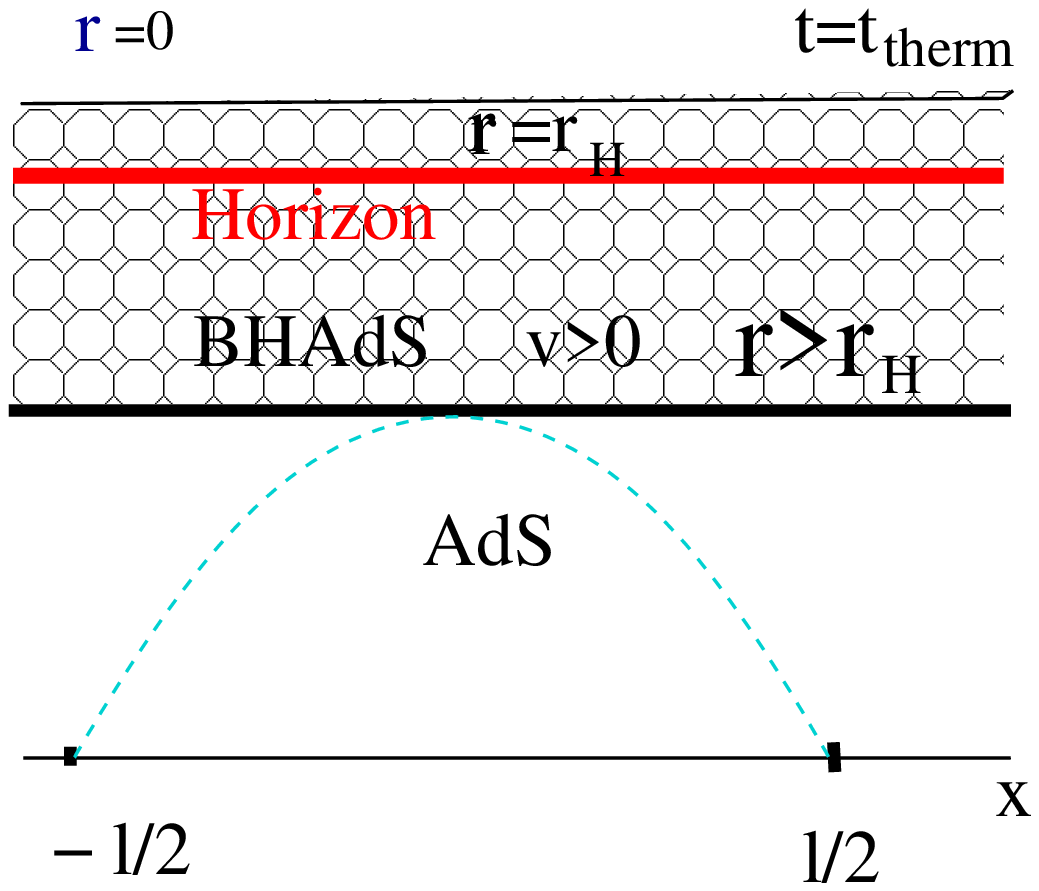}C.\,\,\,\,\,
\includegraphics[width=30mm]{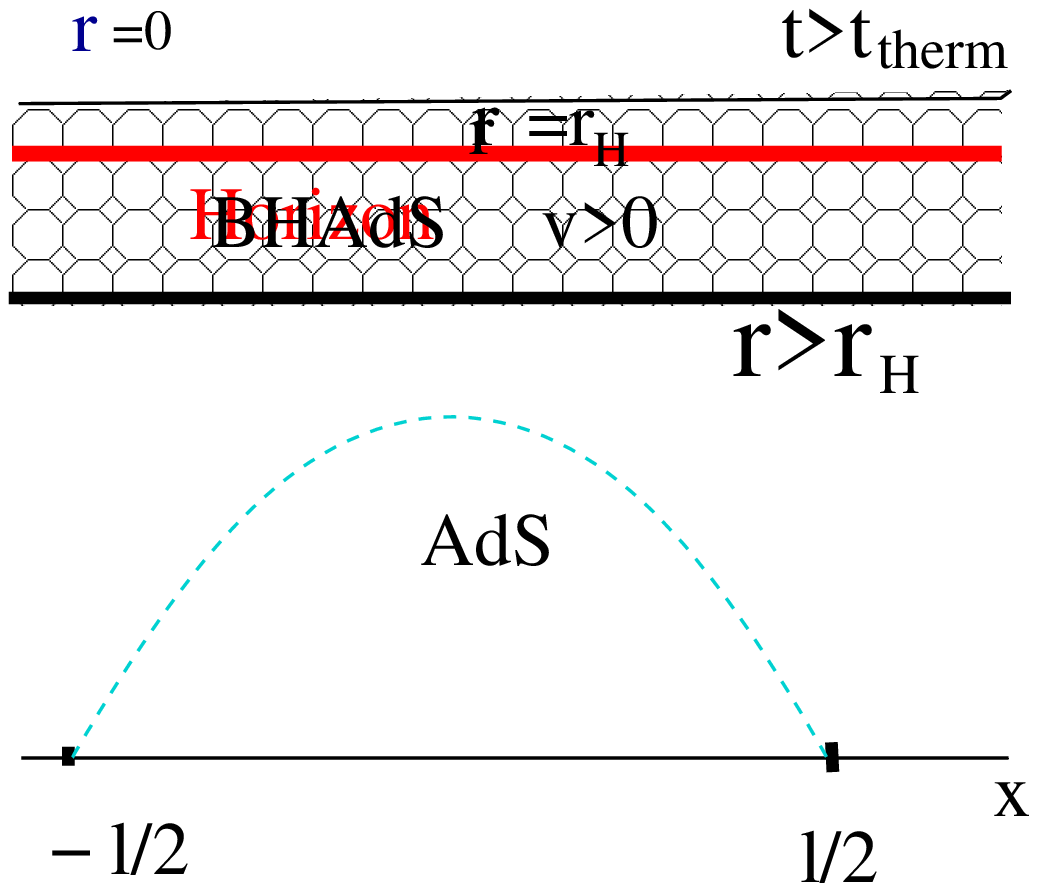}D.
\end{center} \caption{Cartoon of the black hole evaporation: A. The  geodesic connected points  $\pm l/2$ is totally
 in the black hole region. B. A partial evaporation of the black hole from the point of view
 of two points  $\pm \vec \ell/2$ correlator, i.e. the  geodesic is partially
 in the black hole region. Ñ. The total evaporation of the black hole from the point of view
 of this correlator, i.e. the  geodesic totally abandons the black hole region. D. The geodesic
is totally in the empty  region in all subsequent momenta of time.}
\label{Fig:radiation}
\end{figure}

In this section we consider the Hawking radiation in the holographic approach by using the AdS-Vaidya metric and show that the Hawking radiation prevents to thermalization at large distances.
In other words,   the Hawking radiation  in AdS produces the    dethermalization  in dual QGP in the infrared region.
To describe the Hawking radiation we use the  AdS-Vaidya  metric with
negative energy (with negative mass function), compare with  \cite{VZF}, where the Hawking
  radiation has been studied for the Minkowski  Vaidya metric  and see \cite{Callan:2012ip} where
  the AdS-Vaidya  with negative mass has been considered.
The process of the   black hole Hawking radiation in this model  corresponds to
\be \label{nm} m(v)=M-M\theta(v), \ee  and is presented in Fig.\ref{Fig:radiation}.

 The evolution of the 2-point correlation function (\ref{cor}) corresponding to the black hole evaporation  process can be written explicitly in the case $d=2$, see (\ref{l4}), (\ref{L4}) below.
 These formulas  are similar to the evolution formulas describing  the black hole formation
\cite{Balasubramanian:2012}.
 The formulas are obtained by combining relations between
 the change of the value of the affine parameter and the value of the change of the coordinate
   $x$ at the different parts of the geodesic. These relations have the form

\bea\nonumber
l&=&l_{AdS,1}+l_{AdS,2}+l_{BHAdS,1}+l_{BHAdS,2}\\
&=&\frac{4}{p_xr_c}\left(r_c-\sqrt{r^2_c-p_x^2}\right) -2p^2_x\ln
\left(\frac{r_t^2(2-r_c^2G_++2\sqrt{F(r_c)})}{r_c^2(2 -r_t^2G_++2
\sqrt{F(r_t)})}\right)\label{l4} \eea
and
\bea L_{ren}&=&\delta
L_{ren}+L_{AdS,1}+L_{AdS,2}+L_{BHAdS,2}+L_{BHAdS,2}\nonumber\\
&=&-4\log(r_c+\sqrt{r_c^2-p_x^2})
+\ln\left(\frac{-(p^2_{x}+1 -E_B^2)+2r_c^2+2\sqrt{D(r_c)}}
                 {-(p^2_{x}+1 -E_B^2)+2r_t^2+2\sqrt{D(r_t)}}\right)\label{L4} \eea
Here $ren$ means the renormalized length (compare with the action renormalization considered in \cite{IAIV}) and the following notations are used
\bea
G_+&=&p_x^2 +1-E_B^2,\,\,\,\,\,G_-=-p_x^2 +1+E_B^2\\
D(r)&=&r^4+(E_B^2-1-p_x^2)r^2+p_x^2\\
F(r)&=&p_x^2r^4-(p_x^2 +1-\,E_B^2p_x^2)r^2+1
\eea
The above formulas correspond to  the case when in the empty $AdS$ space
the geodesic has zero energy   (the case of nonzero energy corresponds to non-equal time correlator),
and the energy in the black space (in $BHAdS$
space) is defined from the refraction condition
\be
\label{j-E}
E_B=-\frac{1}{2r_c^2}
\sqrt{r_c^2 -p_x^2}.
\ee
Here  $r_c$ is the coordinate of the crossing point,
$p_x$  is the angular momentum that is the integral of motion and
 has no a junction under crossing the geodesic
(as opposed to the energy, that has according to   (\ref{j-E}) a junction under a crossing of the shell),
$r_t$ is the turning point of the geodesic. There is  a relation
\bea
\label{t-p}
r_{t\,\pm}^{2}&=&\frac{(1 +p_x^2-E_B^2)\pm \sqrt{(1 +p_x^2-E_B^2)^2-4p_x^2}}{2}\eea

Let us clarify the meaning of (\ref{l4}), (\ref{L4}), (\ref{j-E}) and (\ref{t-p}).
Here $r_c$ is a free parameter that specifies the position of the shell, $p_x$  is a parameter that does not change
in shell moving, $E_B$ is given by (\ref{j-E}), and $r_{t}$
 given by formula (\ref{t-p}). Therefore, under  $r_c$ and $p_x$ fixed, formulas (\ref{l4}) and (\ref{L4})
give the relation between renormalized geodesic length  $L_{ren}$ and the distance $l$.

\begin{figure}[h!]
\begin{center}
\includegraphics[width=30mm]{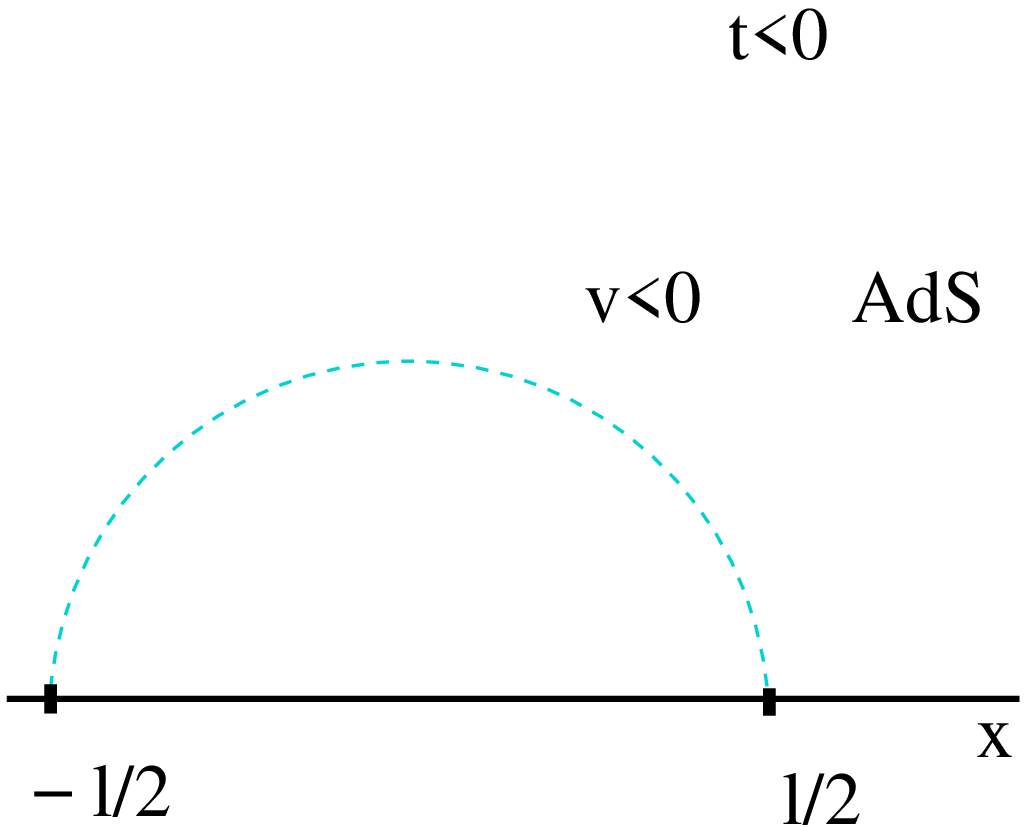}A.\,\,\,\,\,
\includegraphics[width=30mm]{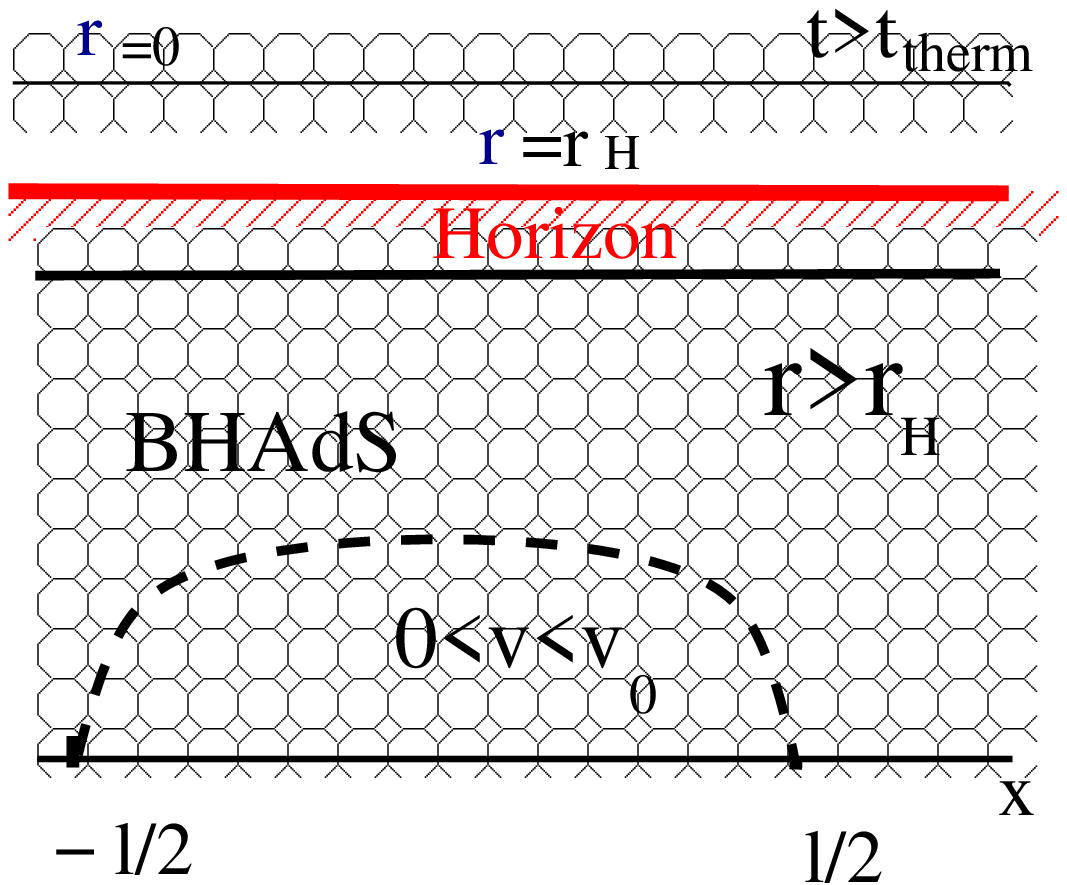}B.\,\,\,\,\,
\includegraphics[width=30mm]{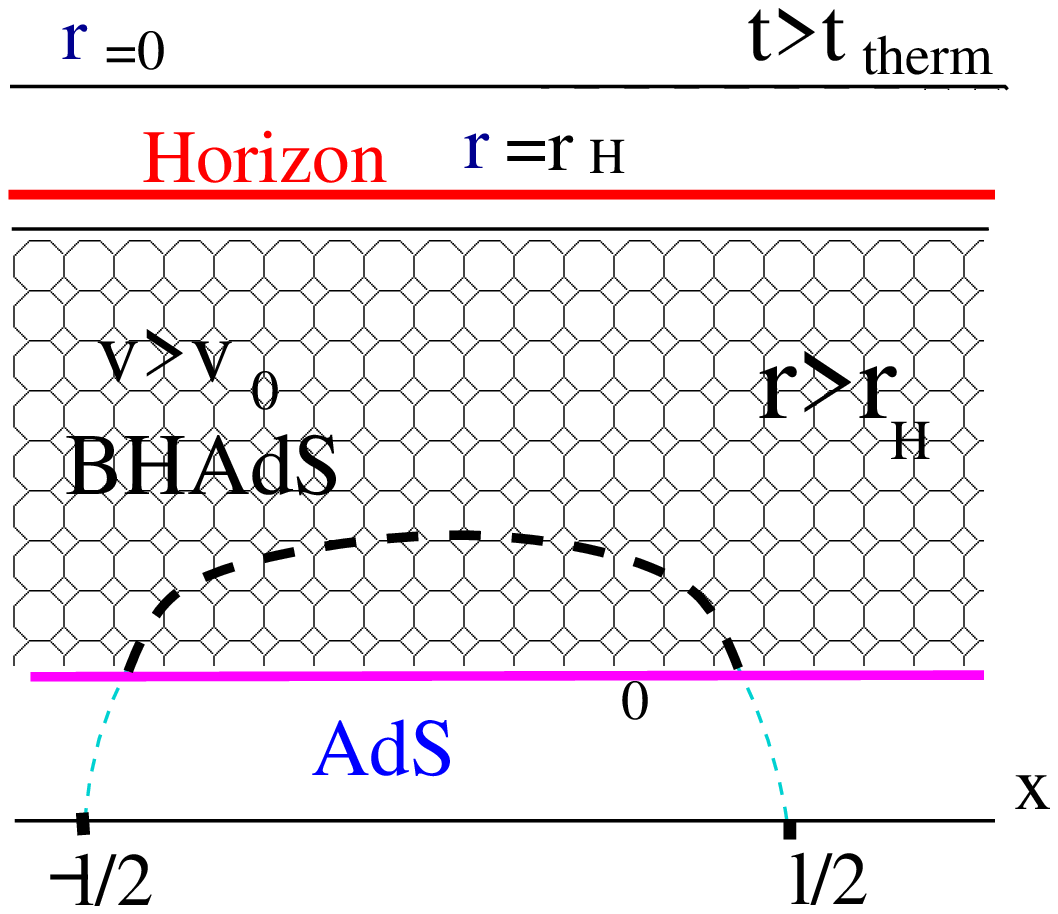}C.\,\,\,\,\,
\includegraphics[width=30mm]{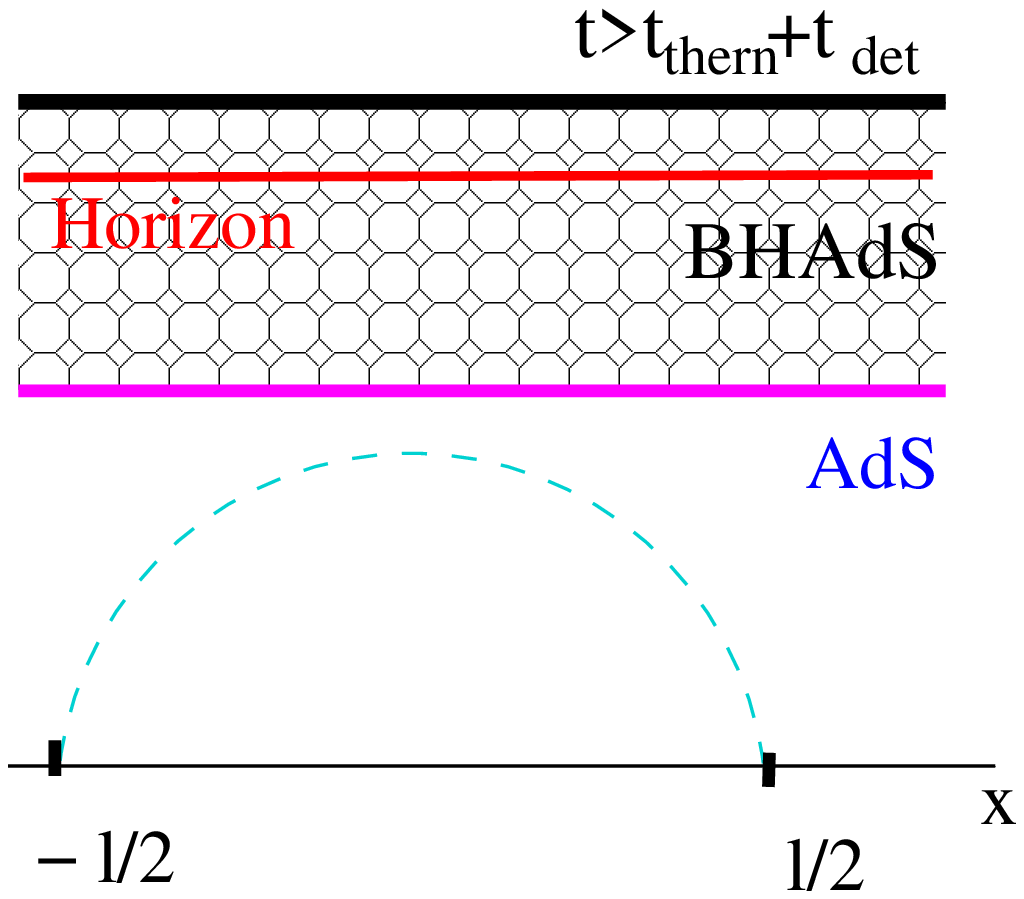}D.
\end{center} \caption{Cartoon of  the black formation  and the subsequent black hole evaporation:
transition from A to B shows  the black hole creation; C shows the black hole evaporation and  D shows the total  black hole
evaporation.} \label{Cartoon-geod-evap}
\end{figure}

The  process of the   black hole creation and subsequent radiation within our model is presented in Fig.
 \ref{Cartoon-geod-evap}. The metric of this process is given by the formula
 (\ref{eq:Vaidya}) with $m(v)$:

\be \label{mv-pm} m(v)=M\theta(v)-M_1\theta(v-v_1), \ee

The case of the total dethermalization corresponds to $M=M_1$ and will be considered in what follows.
If the distance  $\ell$  is much less then
$v_1$, then at the given moment of time the geodesic cross  the shell no more than 2 times. This case is schematically presented in Fig.\ref{Cartoon-geod-evap}.

As we can see from the figures, for  $0<t<v_1$
the length of the geodesic connected the points $(\pm  l/2,t_0,z_0)$
is given by formula  from \cite{Balasubramanian:2012}. We assume that here $t_{therm}<v_1$. Then the  length of the geodesic does not change in the interval   $t_{therm}<t<v_1$, Fig.\ref{Cartoon-geod-evap}B.
Starting from  $v_1=t$
the geodesic crosses the shell with the negative mass (the magenta line in Fig.\ref{Cartoon-geod-evap}. C) at 2 points,
and the  length of the geodesic is given by  formulas (\ref{l4}) and (\ref{L4}) .
 As it can be is seen from Fig.\ref{Cartoon-geod-evap}. D, starting from  $t\geq v_1+t_{det}$
the geodesic related the points $(\pm  l/2,t_0,z_0)$
occurs totally outside the black hole and the geodesic length
 will be given by formula
$\delta {\cal L} = 2 \ln(\ell/2)$.

\begin{figure}[h!]
\begin{center}
\includegraphics[width=30mm]{charts1ggh}A.\,\,\,\,\,
\includegraphics[width=30mm]{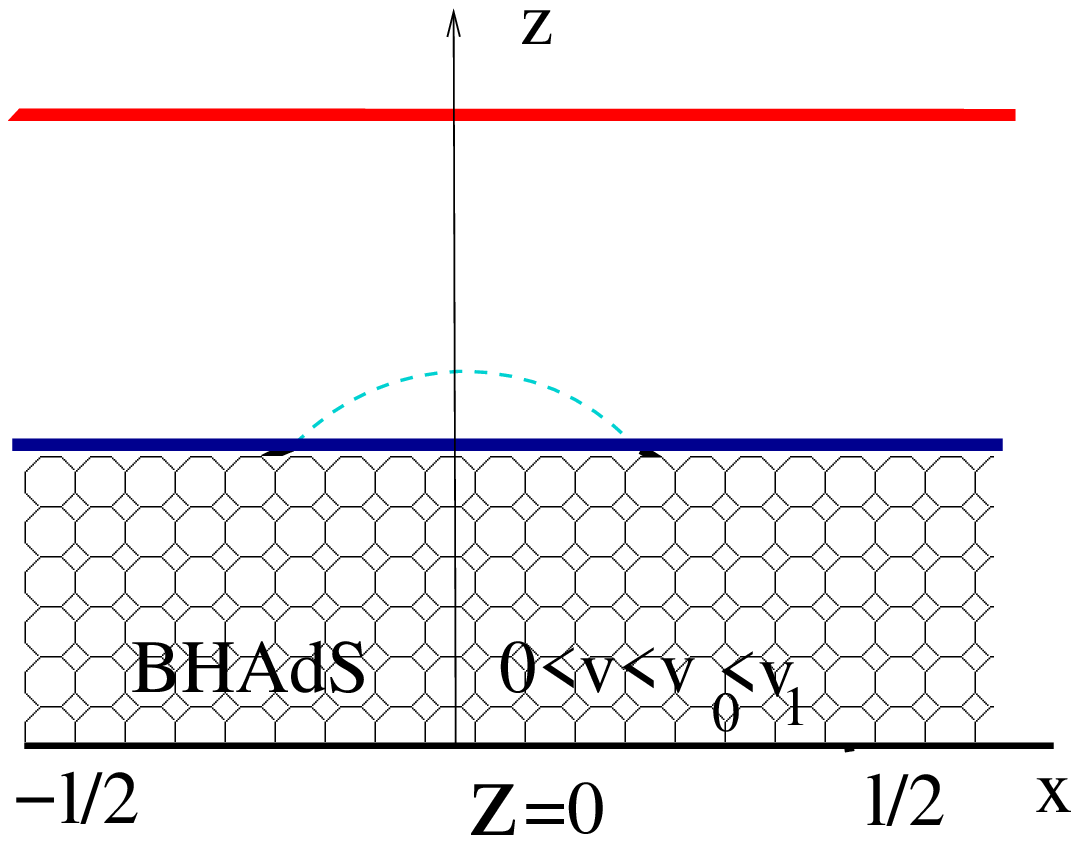}B.\,\,\,\,\,
\includegraphics[width=30mm]{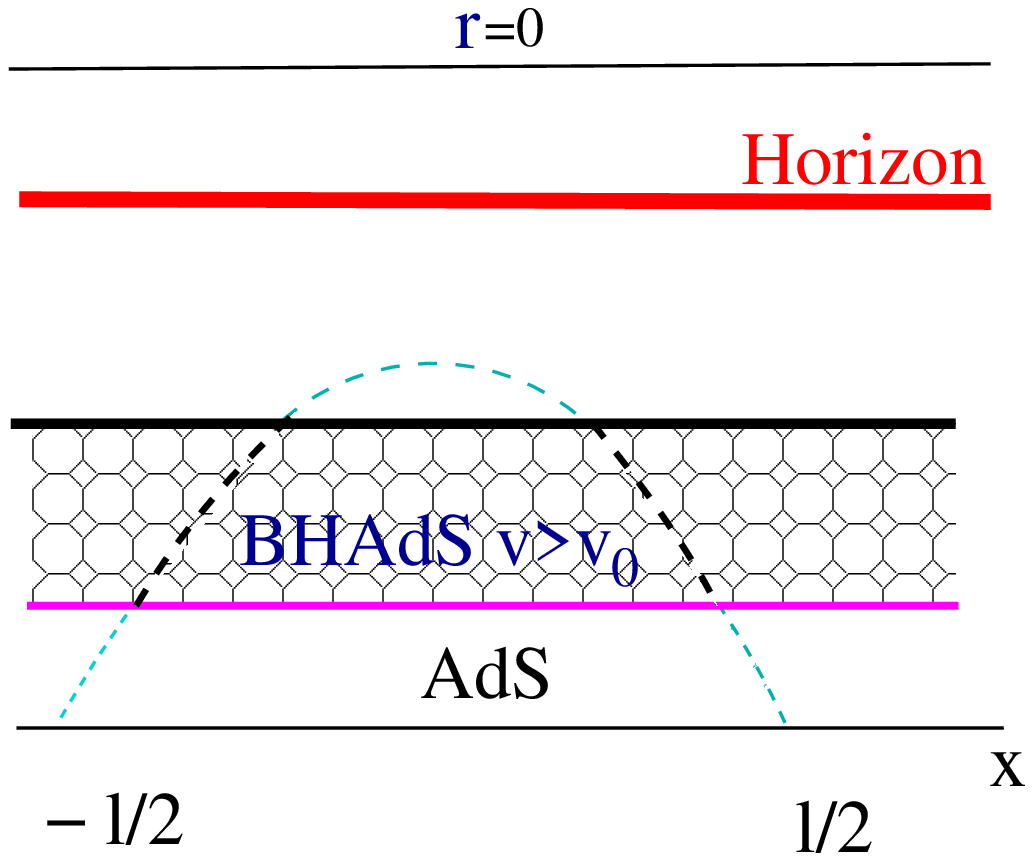}C.\,\,\,\,\,
\includegraphics[width=30mm]{charts12h}D.
\end{center} \caption{Thermalization process does not occur  for a long distance correlations:
B. and C.  show that the black hole does not have  enough time to form, since the
evaporation process has already started; A è D as in Fig.2} \label{Cartoon-geod-evap-large-l}
\end{figure}

For large  $\ell$
the geodesic cross the shell at four  points, see Fig.\ref{Cartoon-geod-evap-large-l},
and the length of the geodesic is given by a more complicated formula, but in any case at large times
the length does not change  and is given by
$\delta {\cal L} = 2 \ln(\ell/2)$,  i.e. the thermalization does not occur.

\section{Bounds on Thermalization and Dethermalization Times}
In this section we establish bounds on thermalization and dethermalization times
in the AdS-Vadiya holographic model.

Let us consider  the AdS-Vadiya metric  (\ref{eq:Vaidya}) with
 \be \label{mv-pm-m} m(v)=M(\theta(v)-\theta(v-v_1)), \ee
where $M>0$ and $v_1$ is larger than the thermalization time, $v_1>t_{ther}$.

Let us suppose that the vector $\vec l$,
characterizing the equal times two points on the boundary, at which
the geodesic states and ends, has only one nonzero component, $l_1\equiv \ell$,
and denote $J_1=J$.
The thermalization time for the correlation function at for these two points is given by
\bea
\tau_{therm}&=&\int  ^{\infty}_{J}\frac{dr}{r^2(1-\frac{M}{r^d})},\label{t-d-m}
\eea
here $J$ is  the first component of the conserved angular momentum related with $\ell$
\bea
\ell &=&2J\int_{J}^{{\infty}}
\frac{dr}{r^2\sqrt{(r^2-J^2)\,(1-\frac{M}{r^d})}}\label{l-d-m}.
\eea
From these formulas we get that
\be
\frac{\tau_{ther}}{\ell}=F(m^2,d),\label{ratio}\ee
where  $m^2=M/J^d\rho^d$ and $F(m^2,d)$ is given by
\be
F(m^2,d)=\frac{\int  ^{\infty}_{1}\frac{d\rho }{\rho^2(1-\frac{m^2}{\rho^d})}}
{2\int_{1}^{\infty}
\frac{d\rho}{\rho ^2\sqrt{(\rho^2-1)\,(1-\frac{m^2}{\rho^d})}}}.\label{F}\ee

\begin{figure}[t!]
\begin{center}
\includegraphics[height=1.5in]{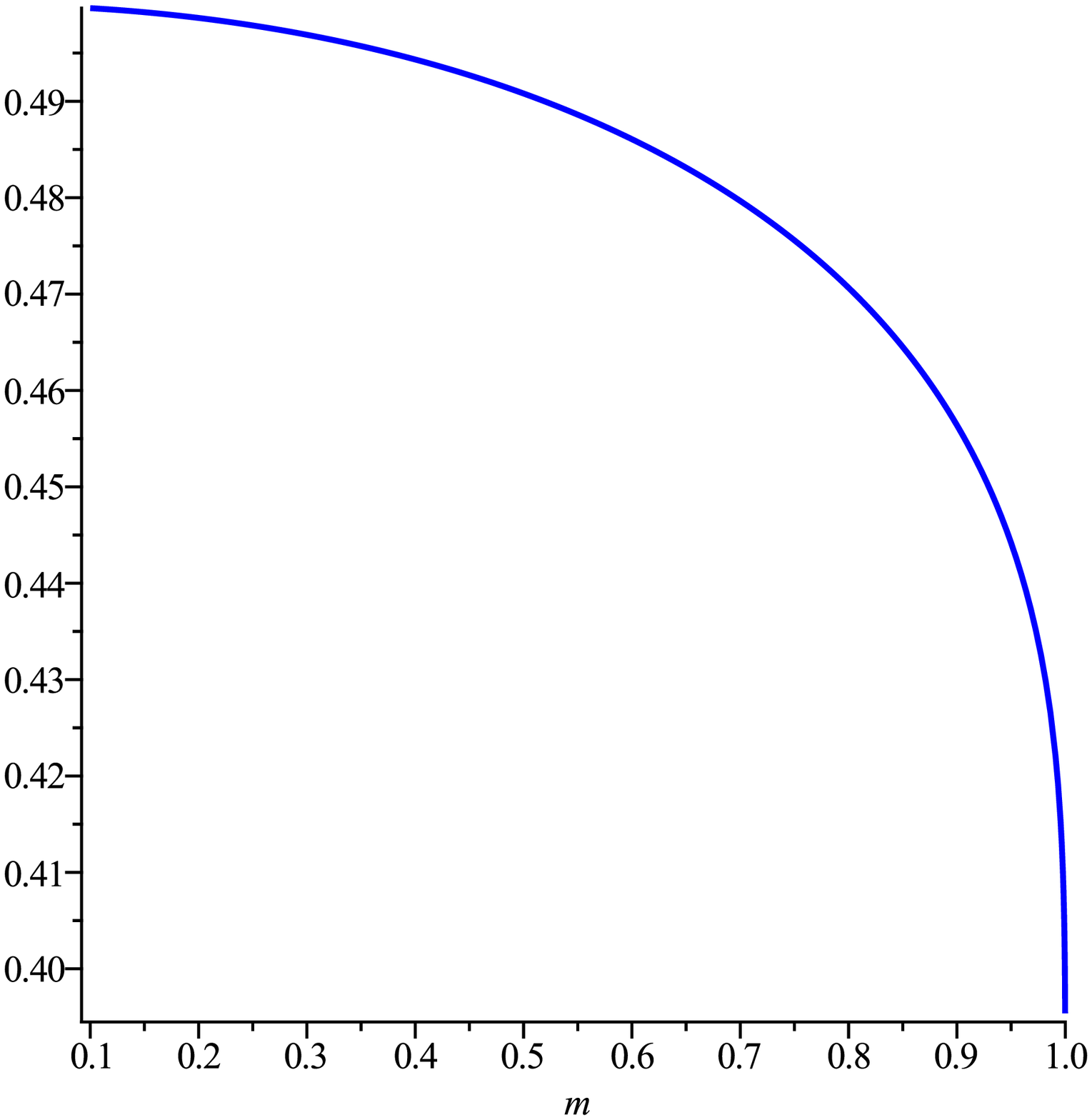}\,\,\,A.\,\,\,\,\,\,\,\,\,
\includegraphics[height=1.5in]{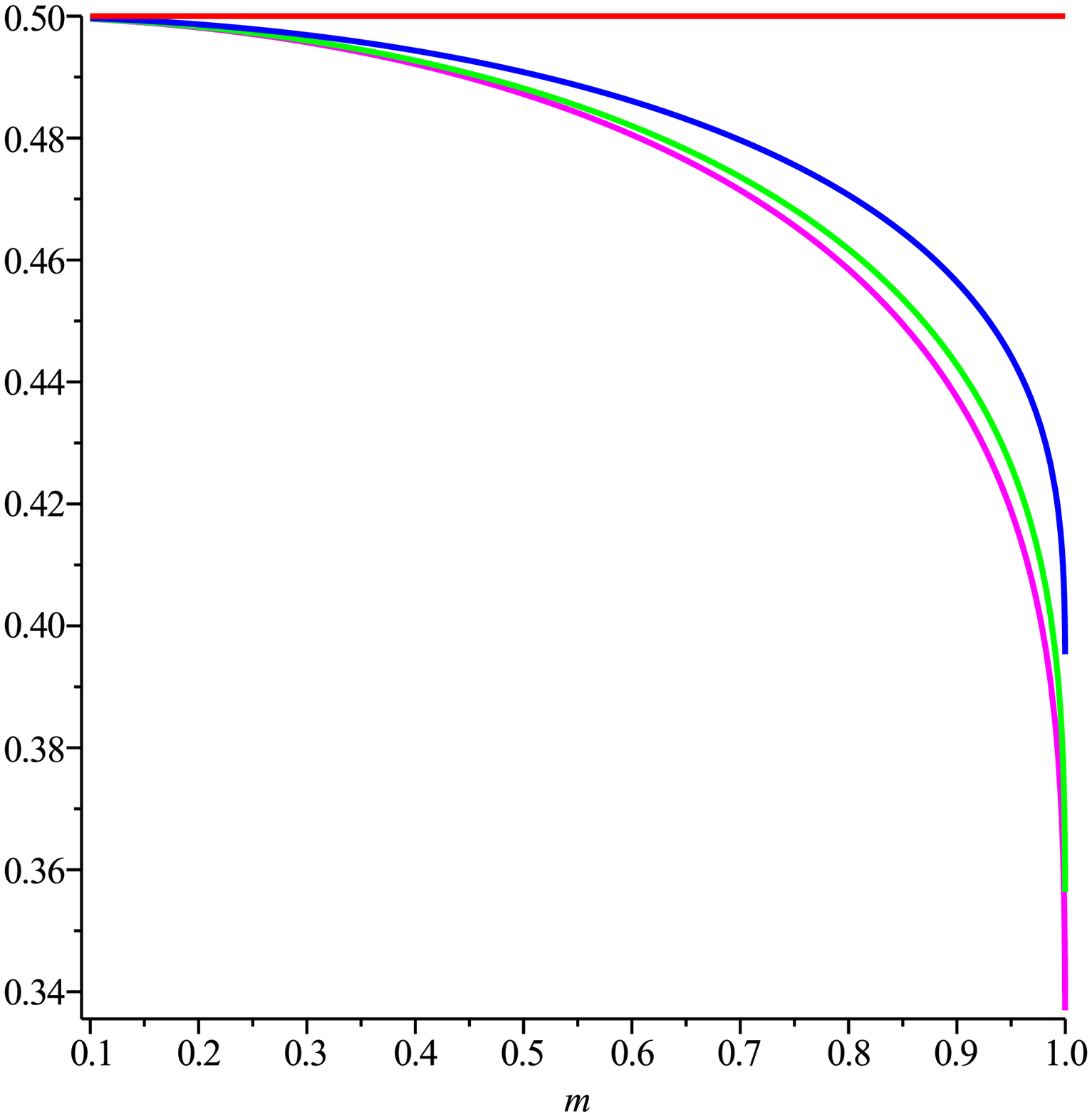}\,\,\,B.
\includegraphics[height=1.5in]{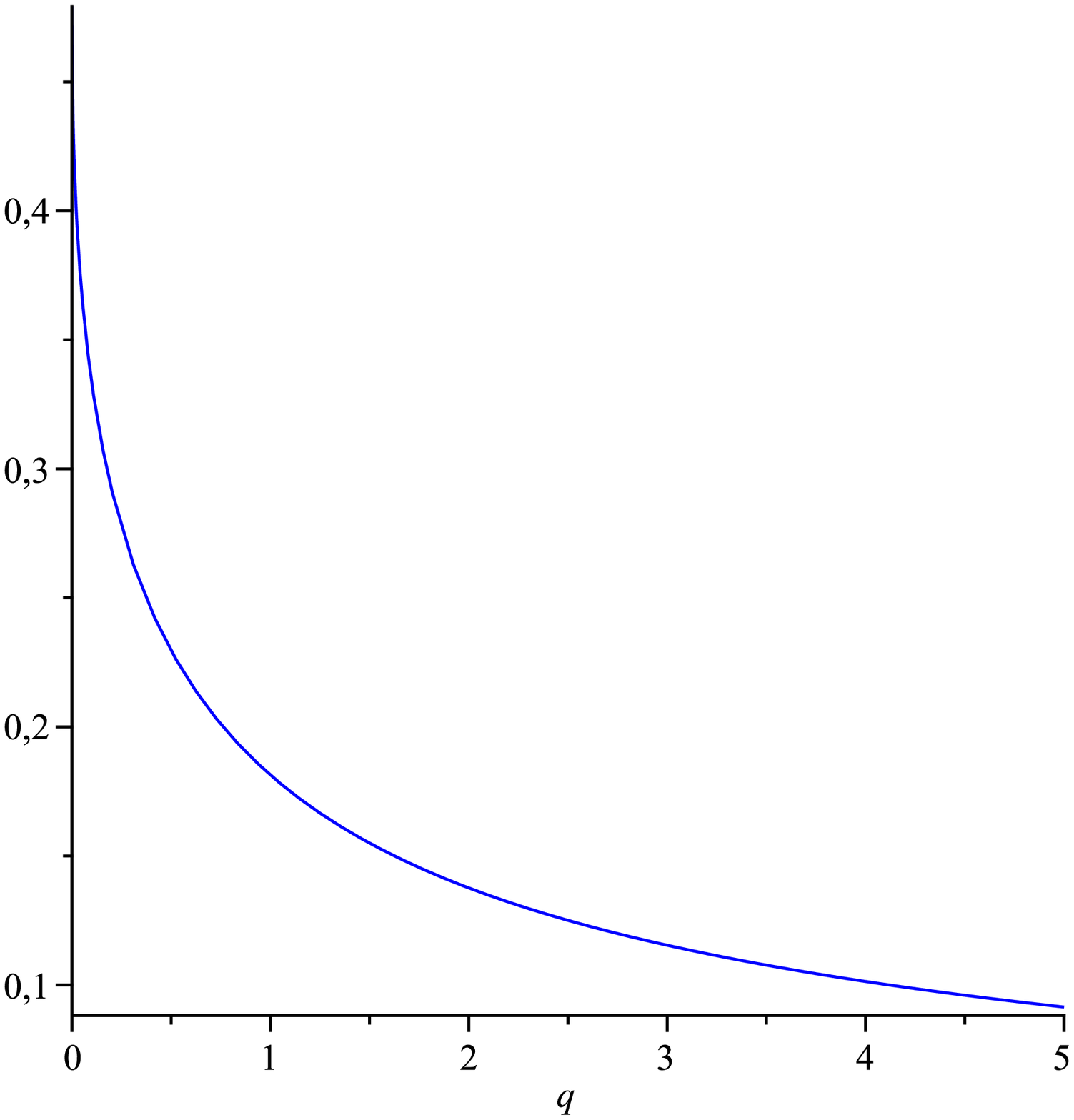}\,\,\,C.\caption
{A. The plot of $F(m^2,4)$ as function of $m$. B. The plot shows the dependence of the below
bound of $F(m^2,d)$ on dimension $d$. d=2 corresponds to  the red line, d=4,6,8 to    blue,  green and  magenta
lines, respectively. C. The plot of $F(m^2,q, 4)$ as function of $q$ and $m^2=0.99$.}\label{Fig:ratio}
  \end{center}
\end{figure}

Now, in the $4$-dimensional Minkowski space, i.e. for $d=4$  for the function
 $F(m^2,4)$ as a function of $m$,  $0<m<1$, we have the bound
\be
0.39\leq F(m^2,4)\leq 0.5,\ee
see Fig.\ref{Fig:ratio}.
Therefore  we obtain  the bound for the thermalization time (d=4)
\be\label{therm}
0.39\leq \frac{\tau_{ther}}{\ell}\leq 0.5,\ee
The similar bounds take place for other $d>2$ \cite{Balasubramanian:2012, AbajoArrastia:2010yt,Callan:2012ip}.

The dethermalization time, under assumption that $v_1>\tau_{th}$ and
for the same points on the boundary, is defined by  the formulas
(\ref{t-d-m}) and (\ref{l-d-m}) with $M=0$.
Since $F(0,d)=1/2$ we have
\be
\frac{\tau_{det}}{\ell}=\frac12\label{ratio-M0}\ee
Note, that (\ref{ratio-M0}) does not depend on the space time dimension $d$.

From (\ref{therm}), (\ref{ratio-M0}) we obtain the following relation between
thermalization and dethermalization times for observables at the same distance:
\be
0.78<\frac{\tau_{ther}}{\tau_{det}}<1\label{ratio}\ee
We see that this ratio is universal and does not depend on the distance between two points till the distance is less then $v_1$.

 Therefore, the minimal  ratio of thermalization time to determalization time, that
can be realized in the $d=4$ AdS-Vadiya model is 0.78. Increasing $d$ one gets a possibility to decrease this ratio, see \ref{Fig:ratio}.B.
As it has been noted in \cite{1205.1548,Elena} involving the nonzero chemical potential
one increases the ratio of $\tau_{therm}$ to $\ell$, and therefore, in our model, this
increases
$\tau_{therm}/\tau_{der}$. One can also try to add by hands an effective locking potential,
for example, the quadratic one. This corresponds to a change $(1-m^2/\rho ^d)\to
(1-m^2/\rho ^d+q\rho ^2)$ in (\ref{F}). This locking potential  decreases the
ratio $F(m,q,4)= \tau_{therm}/\tau_{der}$, see Fig.\ref{Fig:ratio}.C.

It is known that the experimental data on heavy ion collisions  determine
the thermalization time as $\tau_{ther}\sim 1$ fm/c and the dethermalization time as
$\tau_{det}\sim 10- 20$ fm/c,  so $\tau_{ther}/\tau_{det}\sim 0.1 - 0.05$ .
It seems at the first sight that it is difficult to explain this data using our model.
This is in fact  so, if one thinks  that thermalization and the dethermalization
have to be happen at the same scale, but it is not so if  the scales of thermalization
and the dethermalization are different. For the thermalization time,
the relevant length scale according \cite{Balasubramanian:2012}
can be taken about $\ell\sim 0.6$ fm, that is the thermal scale $l\sim \hbar/T$ for the temperature value
$T\sim 300-400 MeV$ at heavy ion collider energies, and  one obtains
the estimate $\tau_{therm}\sim 0.3 $fm/c, which is smaller then the
experimental data.

One can fit better
the experimental data using the scale $l\sim 2$ fm.
One of possible explanations of this scale is  a classical estimation of the
distance between the nucleons  inside the nucleus. One gets this estimation
by taking into account that the radius of the nucleus of Pb is about $r_{Pb}\approx $7 fm,
and in the sphere with this radius one can pack  $208$ (A=208 for Pb) balls with radius
 \be r_n=\,^3\sqrt{\frac{\eta _{K}}{208}\,}\, r_{Pb}\approx 1.07 \,{\mbox {fm}}.\ee
 Here $\eta _{K}$ is the Kepler number $\eta _{K}=\pi/\sqrt{18}\approx 0.74$.
From this consideration it seems natural to take $l=2r_n\sim 2$ fm as a typical scale of the thermalization.
In this case one has $\tau _{therm}\sim$ 1 fm/c.

For an estimation of the dethermalization time,  one can take as the typical scale  the size of the nucleus,
i.e. $l_{det}\sim 2r_{Pb}\sim 14$ fm. Then one gets the dethermalization time
$\tau_{det}\sim 7$fm/c. Note that this estimation gets a low bound, since in the model we have the free
 parameter $v_1$. For the ratio one gets
\be
\frac{\tau_{ther}}{\tau_{det}}
=\frac{\tau_{ther}}{0.5\cdot l_{ther}}\cdot \frac{l_{ther}}{l_{det}}=
0.39\cdot\frac{2}{14}
\approx 0.056,
\ee
which is in agreement with the experimental data.

\newpage

\section{Conclusion}

  In this paper we have used the holographic approach to
study the processes of thermalization and dethermalization in
strongly coupled field theories relevant to the QGP formation in
the heavy ion collisions.

We show that the evaporation of black hole due to the Hawking radiation,
which is modeled by the AdS-Vaidya metric with negative energy,  leads to an interesting phenomena 
in the dual theory -- thermalization is possible only at small distances and impossible in the infrared region.

We have shown that for  the ratio of the thermalization to the dethermalization time, considering at the
the same length scale, has the universal bound and does not depend on the scale.
The assumption that  thermalization and dethermalization take place at the same scale  does not fit well
enough with experimental data.
However it is more reasonable  to  consider   thermalization and dethermalization  at
 different length scales, since  thermalization is local, while  dethermalization 
 is more suitable to relate with largest  space scale of the system under consideration. 
 It is obtained that  the ratio of thermalization time to  dethermalization time is equal 
 to 0.05 that is  consistent with experimental data.

{\bf Acknowledgments.}
The present work is partially supported by the following grants:
RFFI 11-01-00894-a (I.A.), NSch-4612.2012.1. (I.A.) and RFBR
11-01-00828-a (I.V), NSch-2928.2012.1 (I.V).

$$\,$$
\newpage
\appendix
\section{ Hawking radiation and nonequal time correlators.}

\begin{figure}[h!]
\begin{center}
\includegraphics[width=100mm]{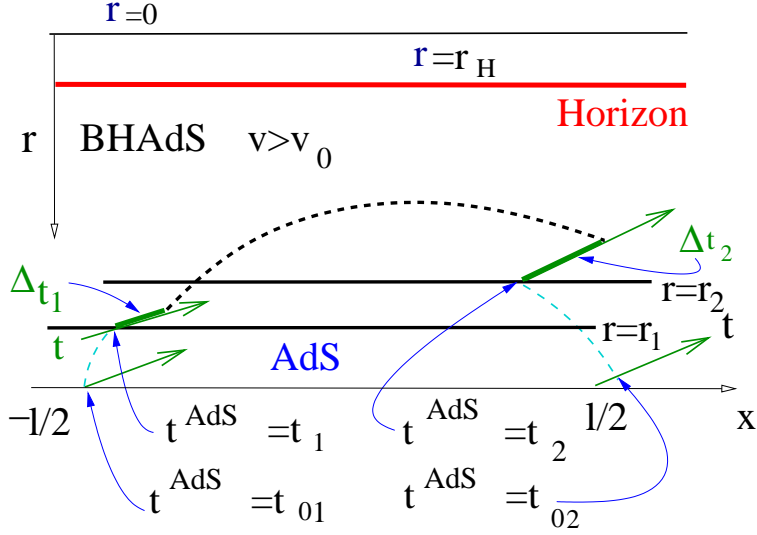}
\end{center}
\caption{A projection of the three dimensional  picture   in coordinates $t,r,x$ of evolution of the shell describing
the process for black hole formation.  The picture contains a  geodesic, related two point on the boundary with different times and
 with non-zero energy
in the empty space.
The green arrows $\rightarrow$ show the time direction. The blue arrows indicate the time coordinates. The down thick line
shows the shell position at the moment  $t^{AdS}=t_1$, and the up thick line shows it at the moment
 $t^{AdS}=t_2$.
A part of the geodesic with nonzero energy in the  empty space is depicted by the dashed line
and a part in the black hole region   by the thick dashed line. geodesic abandon the plane   $t=const$.
The tick green segments show the time junctions under transition from
$AdS$ to $BHAdS$. }
\label{Nmm}
\end{figure}

Considerations in Sect.3 concern to the case
$E=0$ in the AdS space. Here we consider geodesic with nonzero energy in the AdS  space. One of such geodesic is  presented in Fig.\ref{Nmm}.
This figure corresponds to the case when
the shell has no time to
reach the horizon.

 In this case we have

\bea\nonumber
\ell&=&l_{AdS,1}+l_{AdS,2}+l_{BHAdS,1}+l_{BHAdS,2}\\
&=&\frac{2p_x}{(p_x^2-E_1^2)}\left(1-\sqrt{1+\frac{E_1^2-p_x^2}{r^2_1}}\right)
+\frac{2p_x}{(p_x^2-E_2^2)}\left(1-\sqrt{1+\frac{E_2^2-p_x^2}{r^2_2}}\right)\\
&-&p^2_x\ln
\left(\frac{(2
-r_1^2G_++2
\sqrt{F(r_1)})(2
-r_2^2G_++2
\sqrt{F(r_2)})}{p_x^4r_1^2r_2^2}\right)
\\
&+&2p^2_x\ln
\left(\frac{(2
-r_1^2G_++2
\sqrt{F(r_t)})}{p_x^2r_t^2}\right)
\eea
\bea
&\,&L_{AdS,1}+L_{AdS,2}+L_{BHAdS,2}+L_{BHAdS,2}+{\mbox {reg.}}\\\nonumber
&=&-2\ln\left(r_1+\sqrt{r_1^2-(p_x^2-E_1^2)}\right)\left(r_2+\sqrt{r_2^2-(p_x^2-E_2^2)}\right)\\\nonumber
&+&\frac12\ln\left(-\frac12G_++r_1^2+
\sqrt{D(r_1)}\right)\left(-\frac12G_++r_2^2+
\sqrt{D(r_2)}\right)\\\nonumber
&-&2\ln\left(-\frac12G_++r_t^2+
\sqrt{D(r_t)}\right)
\eea

and
\bea
&\,&T_{AdS,1}+\Delta t_1+T_{BHAdS,2}+T_{BHAdS,2}+\Delta t_2+ T_{AdS,2}\\\nonumber
&=&\frac{2E_1}{(p_x^2-E_1^2)}\left(1-\sqrt{1+\frac{E_1^2-p_x^2}{r^2_1}}\right)
-\frac{1}{2}\ln \frac{|r_1-1|}{r_1+1}
-\frac1{r_1}\\
&+&\frac{2E_2}{(p_x^2-E_2^2)}\left(1-\sqrt{1+\frac{E_2^2-p_x^2}{r^2_2}}\right)
+\frac1{r_2} +\frac{1}{2}\ln \frac{|r_2-1|}{r_2+1}\nonumber\\
\nonumber&-&\frac{E}{2E_B}\arctanh
\left(\frac12\,\frac{G_-r_1^2-G_+ }
{\sqrt{D(r_1)}}\right)-\frac{E}{2E_B}\arctanh
\left(\frac12\,\frac{G_-r_2^2-G_+ }
{\sqrt{D(r_2)}}\right)
\\
\nonumber&-&\frac{E}{E_B}\arctanh
\left(\frac12\,\frac{G_-r_t^2-G_+ }
{\sqrt{D(r_t)}}\right)
\eea
\bea E_B&=&E_1-\frac{E_{1}}{2r_1^2}-\frac{1}{2r_1^2}
\sqrt{r_1^2 +E_1^2-p_x^2}\\
E_B&=&E_2-\frac{E_{2}}{2r_2^2}-\frac{1}{2r_2^2}
\sqrt{r_2^2 +E_2^2-p_x^2}\eea

The turning point is now in  BHAdS and it is satisfied
\bea
\label{t-pm}
r_{t\,\pm}^{2}=\frac{(1 +p_x^2-E_B^2)\pm \sqrt{(1 +p_x^2-E_B^2)^2-4p_x^2}}{2}\eea

Here we use
 \bea
\Delta t_1=\Delta_{AdS\to BHAdS}t(r_1)\equiv t^{BHAdS}-t^{AdS}&=&
-\frac{1}{2}\ln \frac{|r_1-1|}{r_1+1}
-\frac1{r_1} \label{Delta1}\\
\Delta t_2=\Delta_{BHAdS\to AdS}t(r_1)\equiv t^{AdS}-t^{BHAdS}&=&
\frac1{r_2} +\frac{1}{2}\ln \frac{|r_2-1|}{r_2+1}
\label{Delta1}\eea

It is clear that if   $r_1=r_2$, then
$\delta t=\Delta t_1+\Delta t_2=0$.


\newpage

\begin{thebibliography}{55}


\bibitem{rhic}
J.~Adams {\it et al.}  [STAR Collaboration],
Nucl.\ Phys.\  A {\bf 757}, 102 (2005) [arXiv:nucl-ex/0501009].

  \bibitem{Gyulassy:2004zy}
  M.~Gyulassy and L.~McLerran,
  Nucl.\ Phys.\  A {\bf 750}, 30 (2005);
  [arXiv:nucl-th/0405013].

\bibitem{fluid2}
 E.~V.~Shuryak,
  Nucl.\ Phys.\  A {\bf 750}, 64 (2005) [arXiv:hep-ph/0405066].


\bibitem{Iancu:2012xa}
  E.~Iancu,
  ``QCD in heavy ion collisions,''
  arXiv:1205.0579 [hep-ph].


  \bibitem{Gelis} 
    F.~Gelis,
  J.\ Phys.\ Conf.\ Ser.\  {\bf 381}, 012021 (2012)
  [arXiv:1110.1544 [hep-ph]].
\bibitem{Muller11} B. Muller and A. Schafer, ''Entropy creation
in relativistic heavy ion collisions'', 1110.2378
\bibitem{Malda} J.~M.~Maldacena,
  Adv.\ Theor.\ Math.\ Phys.\  {\bf 2}, 231-252 (1998) [hep-th/9711200].

\bibitem{GKP} S.~S.~Gubser, I.~R.~Klebanov, A.~M.~Polyakov,
  Phys.\ Lett.\  {\bf B428}, 105-114 (1998) [hep-th/9802109].

 \bibitem{Witten}
  E.~Witten,
  Adv.\ Theor.\ Math.\ Phys.\  {\bf 2}, 253-291 (1998) [hep-th/9802150].


%


\bibitem{EQ-QGP}
 J.~Casalderrey-Solana, H.~Liu, D.~Mateos, K.~Rajagopal and U.~A.~Wiedemann,
  arXiv:1101.0618 [hep-th].


  \bibitem{Gubser} S.S. Gubser, S.S. Pufu and A. Yarom,
Phys.Rev., D 78 (2008) 066014; arXiv: 0805.1551\\
 S.S. Gubser, S.S. Pufu and A. Yarom
\emph{JHEP}11(2009)050; arXiv:0902.4062 .

  \bibitem{Albacete:2008vs}
  J.~L.~Albacete, Y.~V.~Kovchegov and A.~Taliotis,
  JHEP {\bf 0807} (2008) 100, arXiv:0805.2927.

\bibitem{Alvarez}
L. Alvarez-Gaume, C. Gomez, A. Sabio Vera, A.
Tavanfar, M. A. Vazquez-Mozo,
    JHEP 0902:009,2009; arXiv:0811.3969
\bibitem{Chesler:2008hg} P.~M.~Chesler and L.~G.~Yaffe,
  Phys.\ Rev.\ Lett.\  {\bf 102} (2009) 211601, 0812.2053.
 \bibitem{Lin:2009pn}
  S.~Lin and E.~Shuryak,
  Phys.\ Rev.\ D {\bf 79} (2009) 124015, arXiv:0902.1508.



   \bibitem{ABGJ}
 I.~Y.~Aref'eva, A.~A.~Bagrov and E.~A.~Guseva,
 JHEP { \bf 0912}, 009 (2009); arXiv:hep-th/0905.1087\\
I.~Y.~Aref'eva, A.~A.~Bagrov and L.~V.~Joukovskaya,
 JHEP {\bf 1003} , 002, (2010), arXiv:hep-th/0909.1294
 \bibitem{Yaffe} P.~M.~Chesler and L.~G.~Yaffe,
  Phys.\ Rev.\ Lett.\  {\bf 106}, 021601 (2011), arXiv:1011.3562.
\bibitem{KirTalio11} 
  E.~Kiritsis and A.~Taliotis,
  JHEP {\bf 1204}, 065 (2012)
  [arXiv:1111.1931 [hep-ph]].
\bibitem{ABP} I.~Y.~Aref'eva, A.~A.~Bagrov and E.~O.~Pozdeeva,
  JHEP {\bf 1205}, 117 (2012), arXiv:1201.6542.

\bibitem{Bhattacharyya:2009uu}
  S.~Bhattacharyya and S.~Minwalla,
  JHEP {\bf 0909}, 034 (2009)
  [arXiv:0904.0464 [hep-th]].
  \bibitem{Bizon:2011gg}
P.~Bizon and A.~Rostworowski, 
   {\em Phys.Rev.Lett.} {\bf 107} (2011) 031102, arXiv:1104.3702.


  \bibitem{Jalmuzna:2011qw}
  J.~Jalmuzna, A.~Rostworowski and P.~Bizon,
  Phys.\ Rev.\ D {\bf 84}, 085021 (2011)
  [arXiv:1108.4539 [gr-qc]].


  \bibitem{Dias:2011ss}
  O.~J.~C.~Dias, G.~T.~Horowitz and J.~E.~Santos,
  Class.\ Quant.\ Grav.\  {\bf 29}, 194002 (2012)
  [arXiv:1109.1825 [hep-th]].

\bibitem{Garfinkle:2011hm}
  D.~Garfinkle and L.~A.~Pando Zayas,
  Phys.\ Rev.\ D {\bf 84}, 066006 (2011);
  [arXiv:1106.2339 [hep-th]].
  \bibitem{Garfinkle:2011tc}
  D.~Garfinkle, L.~A.~Pando Zayas and D.~Reichmann,
  JHEP {\bf 1202}, 119 (2012),
  arXiv:1110.5823 [hep-th].


 \bibitem{Balasubramanian:2012}
  V.~Balasubramanian {\it et al.},
  Phys.\ Rev.\ Lett.\  {\bf 106}, 191601 (2011)
  arXiv:1012.4753;
  \bibitem{Balasubramanian:2011ur}
  V.~Balasubramanian {\it et al.},
  Phys.\ Rev.\ D {\bf 84}, 026010 (2011)
  arXiv:1103.2683.
\bibitem{AbajoArrastia:2010yt}
  J.~Abajo-Arrastia, J.~Aparicio and E.~Lopez,
   JHEP 1011, 149 (2010).
 arXiv:1006.4090 [hep-th]
  \bibitem{Callan:2012ip}
  R.~Callan, J.~-Y.~He and M.~Headrick,
  JHEP {\bf 1206}, 081 (2012), arXiv:1204.2309.
.

  \bibitem{1205.1548}
  D.~Galante and M.~Schvellinger,
  JHEP {\bf 1207}, 096 (2012)
  [arXiv:1205.1548 [hep-th]].
  \bibitem{Elena}   E.~Caceres and A.~Kundu,
  JHEP {\bf 1209}, 055 (2012)
  [arXiv:1205.2354 [hep-th]].





  \bibitem{Keski}
U.~H.~Danielsson, E.~Keski-Vakkuri and M.~Kruczenski,
  Nucl.\ Phys.\  B {\bf 563}, 279 (1999).
  [arXiv:hep-th/9905227].





   \bibitem{Hubeny:2007xt}
  V.~E.~Hubeny, M.~Rangamani, T.~Takayanagi,
  JHEP {\bf 0707 } (2007)  062, arXiv:0705.0016.
\bibitem{VZF} V. P. Frolov, I. V. Volovich, V. A. Zagrebnov,
     Theor.Math.Phys., 29  (1976) 1012.





  \bibitem{Chesler:2011ds}
  P.~M.~Chesler and D.~Teaney,
  arXiv:1112.6196 [hep-th].

\bibitem{Balasubramanian:1999zv}
  V.~Balasubramanian and S.~F.~Ross,
  Phys.\ Rev.\  D {\bf 61} (2000) 044007, hep-th/9906226.
   \bibitem{IA-cat} I.Ya.~Aref'eva,
Physics of Particles and Nuclei, 41 (2010), 835, arXiv: 0912.5481.



\bibitem{Son}
D.~T.~Son and A.~O.~Starinets,
  JHEP {\bf 0209}, 042 (2002),
  {\tt hep-th/0205051}.
  \bibitem{IAIV}
  I.~Ya.~Aref'eva and I.~V.~Volovich,
  Phys.\ Lett.\ B {\bf 433}, 49 (1998), hep-th/9804182.


\end{thebibliography}
\end{document}